\documentclass[%
 reprint,
superscriptaddress,
nobibnotes,
 amsmath,amssymb,
 aps,
prl,
]{revtex4-2}
\usepackage{svg}
\usepackage{graphicx}
\usepackage{dcolumn}
\usepackage{bm}
\usepackage{bbold}
\usepackage{multirow}
\usepackage[english]{babel}
\usepackage{natbib}
\usepackage{cases}
\usepackage{orcidlink}
\usepackage{lipsum}
\usepackage{dsfont}
\usepackage{lineno}
\begin{document}

\newcommand{\ag}[1]{{\small\bf \textcolor{red}{[AG: #1]}}}
\newcommand{\jr}[1]{{\small\bf \textcolor{blue}{[JR: #1]}}}
\newcommand{\rev}[1]{\textcolor{blue}{#1}}
\newcommand{\rmd}{{\rm d}}
\newcommand{\bphi}{\bar\phi}
\newcommand{\lt}{\left}
\newcommand{\rt}{\right}
\newcommand{\eL}{\mathcal{L}}
\newcommand{\x}{\bm x}
\newcommand{\X}{\bm X}
\newcommand{\y}{\bm y}
\newcommand{\Y}{\bm Y}
\newcommand{\z}{\bm z}
\newcommand{\R}{\bm R}
\newcommand{\DD}{\mathbb{D}}
\newcommand{\cD}{\mathfrak{D}}
\renewcommand{\P}{\bm P}
\newcommand{\brho}{\bm \rho}
\newcommand{\ov}{\mathcal{O}_V}
\newcommand{\bov}{\bar{\mathcal{O}}_V}
\newcommand{\Lam}{\bm \Lambda}
\newcommand{\eLi}{\eL_{\rm{int}}}
\renewcommand{\k}{\bm k}
\newcommand{\en}{\epsilon}
\newcommand{\F}{\bm F}
\newcommand{\E}{\bm E}
\newcommand{\D}{\bm{\mathrm{D}}}
\newcommand{\rmD}{\mathrm{D}}
\newcommand{\Db}{\overline{\bm{\mathrm{D}}}}
\newcommand{\G}{\mathcal{G}}
\newcommand{\bt}{\bar{t}}
\newcommand{\Ray}{\mathcal{R}}
\renewcommand{\v}{\bm v}
\newcommand{\setval}{
    \fmfset{wiggly_len}{2 mm}
    \fmfset{arrow_len}{3mm}
    \fmfset{arrow_ang}{13}
    \fmfset{dash_len}{2mm}
    \fmfpen{0.15mm}
    \fmfset{dot_size}{1.5thick}
    \fmfcmd{
        style_def Double expr p =
        pickup pencircle scaled 0.25mm; 
        draw_dbl_plain p;
        pickup pencircle scaled 0.15mm; 
        enddef;
    }
    
    \fmfcmd{
        style_def Double_wiggly expr p =
        pickup pencircle scaled 0.25mm; 
        draw_dbl_wiggly p;
        pickup pencircle scaled 0.15mm; 
        enddef;
    }
    \fmfcmd{%
    style_def Double_arrow expr p =
        pickup pencircle scaled 0.25mm;
        draw_dbl_plain_arrow p;
        pickup pencircle scaled 0.15mm;
    enddef;
}
    \fmfcmd{
        style_def response expr p =
        pickup pencircle scaled .8 pt;
        idraw ("plain", subpath (0,length(p)/2) of p);
        idraw ("wiggly", subpath (length(p)/2,length(p)) of p);
        enddef;
    }
    \fmfcmd{
        style_def retarded expr p =
        pickup pencircle scaled .8pt;
        idraw ("plain_arrow", subpath (0,length(p)/2) of p);
        idraw ("wiggly", subpath (length(p)/2,length(p)) of p);
        enddef;
    }
    \fmfcmd{
        style_def advanced expr p =
        pickup pencircle scaled .8pt;
        idraw ("wiggly", subpath (0,length(p)/2) of p);
        idraw ("plain_arrow", subpath (length(p)/2,length(p)) of p);
        enddef;
    }
    \fmfcmd{
        style_def keldish expr p =
        pickup pencircle scaled .8pt;
        draw_plain_arrow p;
        pickup pencircle scaled .8 pt;
        enddef;
    }
    \fmfcmd{
        style_def Doubleresponse expr p =
        idraw ("Double", subpath (0,length(p)/2) of p);
        idraw ("Double_wiggly", subpath (length(p)/2,length(p)) of p);
        enddef;
    }
    \fmfcmd{
        style_def Double_retarded expr p =
        idraw ("Double_arrow", subpath (0,length(p)/2) of p);
        idraw ("Double_wiggly", subpath (length(p)/2,length(p)) of p);
        enddef;
    }
    \fmfcmd{
        style_def Double_advanced expr p =
    idraw ("Double_wiggly", subpath (0,length(p)/2) of p);
    idraw ("Double_arrow",  subpath (length(p)/2,length(p)) of p);
    enddef;
    }
     \fmfcmd{
        style_def Double_keldish expr p =
        pickup pencircle scaled 0.25mm;
        draw_dbl_plain_arrow p;
        pickup pencircle scaled 0.15mm;
        enddef;
    }
    
}

\preprint{APS/123-QED}

\title{Self-propulsion of a polaron with an oscillating coupling to its quantum bath}
%

\author{Jacopo Romano}
\affiliation{SISSA --- International School for Advanced Studies and INFN, via Bonomea 265, 34136 Trieste, Italy}
\author{Andrea Gambassi}%
\affiliation{SISSA --- International School for Advanced Studies and INFN, via Bonomea 265, 34136 Trieste, Italy}

\date{\today}

\begin{abstract}
   %
   Motivated by the quest for active quantum matter, we investigate the dynamics of an impurity immersed in a quantum gas --- a polaron --- whose coupling to the surrounding medium is periodically modulated in time, alternating in sign. By integrating out the bath degrees of freedom, we derive an effective velocity-dependent drag force acting on the impurity. Above a critical modulation frequency, the corresponding drag coefficient becomes negative at low velocities, signaling the onset of self-propulsion. In the classical limit, we characterize this transition as a function of the modulation frequency and the bath chemical potential. We then compute the leading-order quantum corrections to the impurity dynamics and show that, while the transition remains robust, it can be suppressed by sufficiently precise measurements of the impurity position.
\end{abstract}

\maketitle



Active particles convert energy into autonomous motion \cite{marchetti_hydrodynamics_2013}, leading to distinctive  non-equilibrium phenomena both at the single-particle \cite{golestanian_propulsion_2005,berg2004coli,di_leonardo_bacterial_2010,romano_anomalous_2026,lagzi_maze_2010} and collective levels \cite{vicsek_novel_1995,toner_long-range_1995,cates_motility-induced_2015,tailleur_statistical_2008,saha_scalar_2020,zwicker_growth_2017,romano_nonreciprocal_2026}. These particles can reach remarkably small length scales, such as nanoscopic self-propelling Janus colloids \cite{paxton_catalytic_2004} and biological nanomotors as small as single proteins or DNA strands \cite{ajaj_molecular_2024,svoboda_direct_1993,mehta_myosin-v_1999,sherman_precisely_2004, omabegho_bipedal_2009}. 
However, even the smallest of these self-propelled agents are still sufficiently large that quantum effects are entirely negligible. Consequently, experimental studies of active matter in the quantum regime remain in their infancy, with only a single proposal reported to date \cite{burgardt_quantum-enabled_2026}.
%

Even from a theoretical perspective it remains unclear how quantum active matter should be realized, 
despite 
several recent 
proposals 
in this direction. 
%
%
At the single-particle level, the relatively limited literature has explored active motion from several distinct perspectives. It has been modeled as a non-unitary random walk~\cite{yamagishi_proposal_2024}, as the biased hopping of a quantum particle coupled to two thermal baths, effectively realizing a quantum motor~\cite{penner_heat--motion_2025}, and as a quantum particle steered along a prescribed classical active trajectory by optical traps~\cite{antonov_engineering_2025}. More recently, quantum analogues of paradigmatic models of classical active particles have been introduced through engineered dissipation\cite{gipouloux_active_2026}.
In many-body settings, activity at the quantum scale has been modeled through non-reciprocal, dissipative interactions between the constituents \cite{takasan_activity-induced_2024, khasseh_active_2025, nadolny_nonreciprocal_2025} or from non-Hermitian terms in effective Hamiltonian \cite{adachi_activity-induced_2022}.
While these models are, in principle, experimentally implementable, their realization generally requires highly controllable synthetic quantum system \cite{barik_quantum_2024,schafer_tools_2020,MULLER20121}  and are so far limited to spatial dimensionalities $d=1$ and 2.
Here we show that these platforms are actually not necessary, as autonomous motion may emerge spontaneously in polaronic systems
%
%
when energy is injected into them by modulating in time 
their interactions.
\begin{figure}
    \centering
    \includegraphics[width=\linewidth]{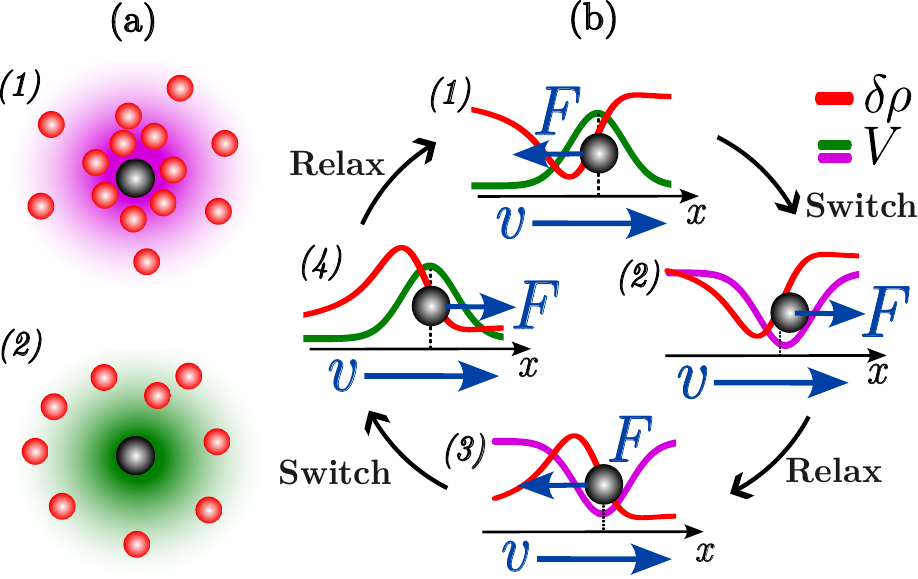}
    \caption{(a) A Bose polaron consists of an impurity (black) immersed in a bath of bosonic particles (red). \textit{(1)} An attractive impurity–bath interaction (purple) leads to a local accumulation of bath particles around the impurity, whereas \textit{(2)} a repulsive interaction (green) produces a local depletion.
    (b) Self-propulsion mechanism: \textit{(1)} A repulsive impurity moving with velocity $v$ creates a wake with density depletion $\delta\rho$ (red line) behind it, which exerts a drag force $\mathbf{F}$. \textit{(2)} When the sign of the interaction potential $V$ (green line) is reversed, the force  exerted by the pre-existing wake changes sign, transiently propelling the impurity forward.  \textit{(3)} As the bath density adapts to the new interaction $V$ (purple line), the force becomes dissipative again. \textit{(4)} A subsequent sign reversal restores the transient propulsive regime. The bath then relaxes back to the dissipative configuration shown in \textit{(1)}, completing the cycle.}
    \label{fig:sketch}
\end{figure}
Polarons are quasi-particles that form when a particle interacts with a quantum bath, often described as a quantum field. The concept was originally introduced 
 to describe electron motion in solids, where lattice phonon excitations act as a bath \cite{landau_electron_1933, pekar1946autolocalization}. In recent years, significant attention has been devoted to polarons realized as impurity atoms in ultracold Bose \cite{chikkatur_suppression_2000,catani_quantum_2012,spethmann_dynamics_2012, scelle_motional_2013} and Fermi gases \cite{massignan_polarons_2014,kohstall_metastability_2012,koschorreck_attractive_2012}. These systems have the remarkable property that the impurity–bath interactions can be tuned via a Feshbach resonance, allowing the realization of both attractive and repulsive polarons \cite{chin_feshbach_2010,schirotzek_observation_2009,kohstall_metastability_2012,koschorreck_attractive_2012}, --- such as those sketched in Fig.~\ref{fig:sketch}(a) --- as well as driven polaron with oscillating interaction strength \cite{vivanco_strongly_2025}. In both cases, impurity–bath interactions renormalize the mass of the impurity \cite{trefzger_energy_2013,nascimbene_collective_2009} 
 and give rise to dissipative effects, including an effective drag that slows down the impurity \cite{astrakharchik_motion_2004,sykes_drag_2009}. 
The physical origin of this drag can be understood as follows. As the impurity moves through the bath, it distorts the surrounding medium, generating a wake of depleted (accumulated) density for repulsive (attractive) interactions. In either case, the wake exerts a backward force on the impurity, giving rise to a drag that opposes its motion, as in Fig.~\ref{fig:sketch}(b\textit{1}). Suppose now that, at some instant, the impurity–bath coupling reverses sign on a timescale shorter than that required for the wake to relax the excess energy. The interaction with the pre-existing wake then temporarily switches from dragging to propelling the impurity, as in Fig.~\ref{fig:sketch}(b\textit{2}). As the wake gradually adapts to the new coupling, the force eventually becomes dissipative again, as shown by Fig.~\ref{fig:sketch}(b\textit{3}).
A subsequent sign reversal of the impurity–bath coupling restores its original sign while causing another transient propulsive regime, as in Fig.~\ref{fig:sketch}(b\textit{4}). The bath then relaxes back to the dissipative
configuration shown in Fig.~\ref{fig:sketch}(b\textit{1}), completing the cycle, which can now be repeated.
In this work, we demonstrate 
that this simple mechanism can be harnessed to induce autonomous motion of the impurity by periodically reversing the sign of its interaction.
In particular, we consider an impurity in a Bose gas \footnote{\label{fn:semiclassical}Since the semiclassical limit considered in this work is insensitive to the underlying particle statistics, the results presented here apply also to Fermi gases.} above the condensation temperature 
and integrate out the bath degrees of freedom perturbatively, leading to an effective drag on the particle and to 
noise.
In order to characterize the features of the resulting effective particle dynamics and highlight the possible emergence of activity, we consider its semiclassical limit.
%
%
We show that, at sufficiently high modulation frequencies, the (speed-dependent) mean drag over one modulation period becomes aligned with the direction of motion, corresponding to a negative drag coefficient for the particle. 
This marks the onset of self-propulsion in a spontaneously selected direction,
which we characterize by determining the critical boundary for this 
transition and by computing the self-propulsion speed for sinusoidal modulations.
We also investigate the features of the wake, showing that in the self-propelling regime the point of maximal field perturbation can \textit{anticipate} rather than follow the particle position, an effect that is key to the self-propulsion. We then explore how quantum fluctuations modify this classical picture. 
%
We show that, while the self-propulsion transition is generically robust, it can be suppressed, at least at intermediate times, by the momentum uncertainty associated with preparing a spatially localized initial state. A similar suppression arises from frequent and sufficiently precise measurements of the impurity position, highlighting a fundamental distinction between classical and quantum self-propulsion.

Finally, we derive an expression for the velocity autocorrelator, highlighting the interplay between stochastic and quantum fluctuations in determining the mean-squared displacement (MSD) of the impurity.

\paragraph*{The model.} 
The 
Hamiltonian $H(t)$ of the system is given by 
$H(t)=H_I+H_B+H_{\rm{int}}(t)$, where $H_I$ and $H_B$ are the Hamiltonians of the isolated impurity and Bose gas, respectively, while $H_{\rm{int}}(t)$ describes the impurity-bath interaction. 
In particular, $H_I=\P^2/(2 m_I)$, where $\P$ and $m_I$ are the momentum and mass of the impurity, respectively. The bath is described as a complex bosonic quantum field $\phi(\x,t)$ with Hamiltonian: $H_B=\int \rmd^dx \big[\frac{\hbar^2}{2m}|\nabla\phi|^2-\mu|\phi|^2\big]$, where $m$ is the mass of the bath particles, $\mu$ the chemical potential and $d$ the spatial dimensionality of the system. 
%
%
We assume the gas to be above the condensation transition, i.e., $\mu<0$, and neglect the interactions between the  bath particles. This allows us to neglect nonlinear contributions $\propto |\phi|^4$ in $H_B$. 
Finally, the bath-impurity interaction is given by $H_{int}(t)=\int \rmd^dx \,V(\x-\R(t),t)|\phi(\x,t)|^2$, where $\R(t)$ is the impurity position and $V(\x,t)=[g(t)/(2\pi a^2)^{d/2}]\exp\lt[-\x^2/(2a^2)\rt]$ 
models a short-range interaction potential with oscillating strength $g(t)$. In particular, we focus below on the case $g(t)=g_0\cos(\omega t)$ but emphasize that a different periodic modulation would lead to the same phenomena, provided that $g(t)$
acquires both positive and negative values during a period. 
%
%
At time $t=0$ we assume that the impurity in the state $|\psi_0\rangle$ is inserted at rest in the bath. 
 The initial density matrix $\brho_0$ of the system is therefore $\brho_0=|\psi_0\rangle\langle\psi_0|\otimes\brho_B$, where $\brho_B$ is the density matrix of a 
 free Bose gas at thermal equilibrium at temperature $T$. 
The resulting non-equilibrium dynamics of the impurity can be derived by integrating out the bath degrees of freedom within 
the Keldysh formalism \cite{kamenev_field_2023}.
Here we report the essential steps of the analysis, while referring to  Ref.~\cite{noauthor_see_nodate} for details. 
Within this formalism, the dynamics is encoded in a generating functional $Z$, constructed from the Hamiltonian $H(t)$ and the initial density matrix $\brho_0$ as a path integral over the bath fields and the impurity coordinates. We perturbatively integrate out the bath, obtaining an effective generating functional  
$Z=\int \mathcal{D}\R_{c,q}\,\rho_0(\R_c^0,\R_q^0)\,e^{i S_A/\hbar}$ for the impurity alone.
Here $\mathcal{D}\R_{c,q}$ denotes integration over the classical and quantum components  $\R_c$ and  $\R_q$, respectively,  of the impurity trajectory,  while $\rho_0(\R_c^0,\R_q^0)$ is their initial distribution 
induced by $\brho_0$. 
We compute the action $S_A$ at the second order in $g(t)$, as the first vanishes \cite{noauthor_see_nodate}. 
In order to highlight the features of the resulting dynamics, we focus below on its semiclassical limit. In this case, one finds 
$S_A=S_0+S_{\rm{diss}}+S_{\rm{fluc}}$, where $S_0=2\hbar\int_{0}^{+\infty}\! \rmd t \, m_I \dot\R_q(t)\cdot\dot \R_c(t)$ 
is the Keldysh action of the free impurity, while $S_{\rm{diss}}$ and $S_{\rm{fluc}}$ describe, respectively, the dissipation and the fluctuations caused by 
the bath. For the dissipation part we have:
\begin{align}
\label{eq:diss_part}
     &S_{\rm{diss}}=2\hbar\int_{0}^{\infty}\!\! \rmd t\, R^i_q(t)\int_{0}^t\!\! \rmd t'E^i(\Delta \R_c(t,t'),t,t'),\\
    &\E(\x,t,t')=\frac{g(t)g(t')}{(2\sqrt{\pi}a)^d}\int \!\rmd^d y\, e^{-\frac{\y^2}{4a^2}}\bm{\nabla} \Lambda(\x+\y,t-t'),\nonumber
 \end{align}
 with $\Delta \R_c(t,t')=\R_c(t)-\R_c(t') = \int_{t'}^t \!\rmd t'' \dot\R_c(t'') $ and $\Lambda(\x,t)=\hbar\,\mathrm{Im}\left[G_R(\x,t)G_K(\x,-t)\right]$, where $G_R$ and $G_K$ are the retarded and Keldysh components of the free boson propagator at high temperature $T$, whose expressions are given in Ref.~\cite{noauthor_see_nodate}.
 Since terms proportional to $2\R_q$ in the Keldysh action have the interpretation of a classical force $\F$, Eq.~\eqref{eq:diss_part} shows the emergence of a drag $\F(\{\dot\R_c\},t)=\int_{0}^t \rmd t'\E(\Delta \R_c(t,t'),t,t')$ on the impurity. 
 This is given by the integral over time, from the initial time up to $t$, of the generalized friction kernel $\E$ calculated at previous particle positions, and thus it depends non-linearly on the past velocity  of the impurity. 
 Note that $\F(\{\mathbf{0}\},t) = \mathbf{0}$ because $G_{R,K}(\x,t)$ are even functions of $\x$.
 %
  \begin{figure*}[th!]
  \includegraphics[width=\textwidth]{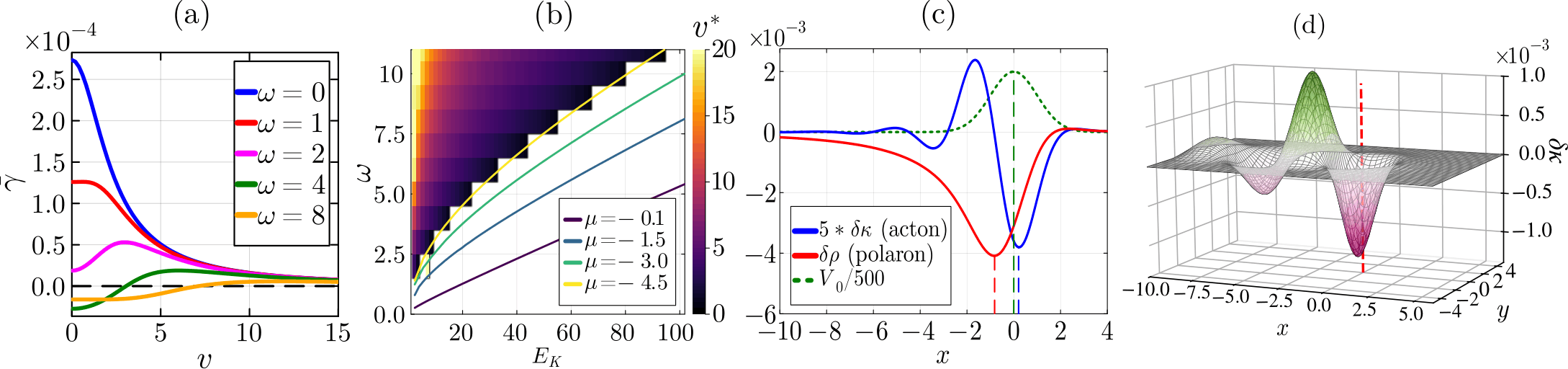}
  \caption{%
  (a) Plot of the average viscosity $\bar\gamma(v)$ for various modulation frequencies $\omega$.
(b) Critical lines $\omega=\omega_c (E_K)$, where $E_K$ is tuned through the interaction width $a$, for various values of the chemical potential $\mu$. For $\omega > \omega_c$ one observes self propulsion.
The density plot shows the self-propulsion speed $v^*$ in the active phase for
$\mu=-4.5$.
(c) Red curve: wake $\delta\rho$ for a repulsive polaron moving at velocity $v=4$ along the $x$ direction. Blue curve: mean wake $\delta\kappa$ of an acton moving in the positive $x$ direction with the same velocity. Both curves are in the co-moving frame.
For the acton, we set $\omega=10$,
corresponding to a self-propulsion speed $v^*=9.05>v$. 
The vertical dashed lines
indicate the location of the minima of the corresponding wakes. The green dotted curve shows
$V_0(x)$, while the associated dashed line marks its maximum.
Both $\delta\kappa$ and $V_0$ are multiplied by a scaling factor for visualization purposes.
(d) Surface plot of $\delta\kappa(\mathbf{x})$ in the $x$--$y$ plane for an
acton moving at the self-propulsion velocity $v=v^*$ along the positive $x$-axis with $\omega=10$. The
red dashed line indicated horizontal location of the center of $V$.
In all panels we set $g_0=T=m=1$ and considered $d=3$. In panels (a), (c), and (d) we also
use $\mu=-1$ and $a=1$. Note the small values of $\bar\gamma$, $\delta\rho$, and $\delta\kappa$,
which originate from a small prefactor $(2\pi)^{-d}$ and from  $\Lambda$ being exponentially suppressed over a large part of the integration domain. 
}
  \label{fig:wideimage}
\end{figure*} 
%
%
 For the fluctuation part we have:
 \begin{align}
 \label{eq:fluc_part}
&S_{\rm{fluc}}=2i\hbar\int_{0}^{\infty}\!\! \!\rmd t\, \rmd t'R^i_q(t)R^i_q(t') \cD^{ij}(\Delta\R_c(t,t'),t,t'),\\
    &\cD^{ij}(\x,t,t')=\frac{g(t)g(t')}{2\sqrt{\pi}a}\!\int \!\rmd^d y\, e^{-\frac{\y^2}{4a^2}}\partial^i\partial^j \Pi(\x+\y,t-t'),\nonumber
 \end{align}
 with $\Pi(\x,t)=-(\hbar^2/4)|G_K(\x,t)|^2$. 
 The matrix $\cD$ describes the variance of the thermal noise induced on the impurity. As for the drag, it depends on the impurity trajectory, resulting in a multiplicative, non-Markovian noise. In the passive case $\omega=0$, the friction kernel and the noise are related by  the fluctuation-dissipation relation (FDR)  $\partial_t \cD^{ij}(\x,t-t')=k_BT\partial^iE^j(\x,t-t')$ for $t>t'$  \cite{noauthor_see_nodate}. This is similar to the FDR derived for a classical overdamped probe in a  medium with dissipative dynamics \cite{basu_dynamics_2022,demery_non-gaussian_2023}.
 %
 %
%
We also note that, due to the structure of $G_{K,R}$ 
\cite{noauthor_see_nodate}, both $\Pi$ and $\Lambda$ grow upon decreasing $\hbar$. 
The friction and noise kernels are characterized by three 
timescales: the driving period $\tau_{\omega} = 2\pi/\omega$ and two internal scales  $\tau_{\mu}= \hbar/|\mu|$ 
and $\tau_{K}$, associated with the kinetic energy of the wake.
In particular, $\tau_{K} = \hbar/E_{K}$, where $E_{K} = h^2/(2 m a^{2})$ 
is the characteristic kinetic energy of a gas particle in the wake 
\footnote{Since the characteristic length scale of the field perturbation is set
by the spatial range $a$ of the potential $V$, its typical momentum is
$p_w = h/a$, corresponding to a kinetic energy
$E_K = p_w^2/(2m) = h^2/(2ma^2)$.}. 
The impurity dynamics introduces an additional timescale $\tau_D$, set by the typical time required for the impurity to change velocity appreciably due to the interaction with the gas. 
This timescale is proportional to $m_I/g_0^2$, 
which controls the relative strength of bath interactions w.r.t.~inertia. In the weak-coupling regime, characterized by small values of $g_0$, 
one expects $\tau_D\gg\tau_K,\;\tau_\mu$. Accordingly, $\dot\R_{c}$ 
%
%
changes slowly w.r.t.~the typical memory timescales of the bath and one can thus approximate $\Delta\R_c(t,t')\simeq \dot\R_c(t)(t-t')$ in Eq.~\eqref{eq:diss_part}. 
This leads to a Markovian drag $\F(\v,t)=\int^{\infty}_{0} \rmd t' \E(\v t', t,t-t')$ 
which depends on the (slow varying) impurity velocity $\v=\dot \R_c(t)$ and on time $t$. 
We write it as $\F(\v,t)=-\gamma(v,t)\v$, introducing the effective  viscosity $\gamma(v,t)$ of the bath. 
We also define the time average over a period of a time dependent quantity $A(t)$ as $\bar A= \tau_\omega^{-1}\int_0^{\tau_\omega}\rmd  t\, A(t)$ and focus on the average drag $\bar \F(\v)$ and viscosity $\bar\gamma(v)$. 
Then, if we also assume $\tau_D\gg\tau_{\omega}$, oscillations of $g(t)$ are fast on the scale of $\tau_D$ and they can be averaged out. As a result, the effective evolution of the particle is governed solely by $\bar\gamma$ and $\bar \F$. Under the same assumption of separation of timescales we define: $\DD(\v,t)=\int_0^{ \infty} \rmd t' \,\cD(\v t',t,t-t')$
and the corresponding time average $\overline{\DD}(\v)$. Then, we can replace $\cD(\Delta\R_c(t,t'),t,t')\simeq  2\Bar\DD(\v)\delta(t-t')$
in Eq.~\eqref{eq:fluc_part}, resulting in a Markovian multiplicative noise on the impurity. Below 
we study $\bar\gamma$ as a function of the model parameters in order to characterize the emergence of self-propulsion, and then assess the impact of quantum and thermal fluctuations on this behavior.

 \paragraph*{Self-propulsion and wake properties.}
 We compute $\bar\gamma(v)$ by numerical integration of $\gamma(v,t)$ 
 and plot it in Fig.~\ref{fig:wideimage}(a) as a function of $v$ for various $\omega$, with fixed $E_K$ and $\mu$, in spatial dimension $d=3$.
In the passive case $\omega=0$, the friction $\gamma$ is a positive function of $v$. 
%
%
%
%
As the modulation frequency $\omega$ increases, however, $\Bar\gamma(v)$ develops first an inflection point and then a local minimum around $v=0$, becoming negative at small $v$ once $\omega$ exceeds a critical threshold $\omega_c$. 
%
%
At larger velocities, instead, the influence of the modulation of the bath on the particle becomes progressively weaker and, as $v$ increases, all curves converge towards that of the case $\omega=0$  with $\Bar\gamma(v)>0$. 
In fact, since faster impurities get farther from their previous location in a given time, they also interact less with the field perturbation produced earlier. Consequently, the drag becomes both weaker and less sensitive to the modulation of $g(t)$. 
%
%
%
%
For $\omega>\omega_c$, $\bar{\gamma}$ changes sign from negative to positive as the impurity velocity $v$ increases. This implies the existence of critical velocity $v^*$, defined by the condition $\bar{\gamma}(v^*)=0$. The medium accelerates impurities with $v<v^*$ while slowing down those with $v>v^*$, driving the particle toward the stable fixed point $v=v^*$. Therefore, the sign change in $\bar\gamma$ marks the onset of self-propulsion in the classical dynamics, with the impurity approaching the asymptotic propulsion speed $v^*$ at long times.
We note that transition from $v^*=0$ at $\omega<\omega_c$ to $v^*>0$ for $\omega>\omega_c$ is continuous, as shown in Fig. \ref{fig:wideimage}, and that the same qualitative behavior discussed above is observed in $d=1$ and $2$.
%

%
%
%

%
In Fig.~\ref{fig:wideimage}(b) we characterize the self-propulsion transition by plotting the critical line $\omega=\omega_c$ as a function of $E_K$ for various values of $\mu$. 
For the largest value of $|\mu|$, we additionally show a color map of the corresponding 
$v^*$ as a function of $E_K$ and $\omega> \omega_c(E_K)$. 
We see that, for fixed $E_K$ (fixed $\omega$) the transition happens at increasing values of $\omega$ (decreasing values of $E_K$) as $\mu$ decreases. 
This is due 
due to the smaller value of the associated memory timescales $\tau_\mu$ and $\tau_K$. 
%

%
%
In order to understand the emergence of the negative drag it is instructive to study in detail the wake, responsible for the active behavior of the impurity.
In particular, we study how the average 
density $\rho(\x,t)=\langle|\phi(\x,t)|^2\rangle$ is perturbed by an impurity moving with a classical trajectory of velocity $\v$ (with $\R_q(t)=0$ and $\R_c(t)=\v t$), by computing its perturbation $\delta \rho(\x,t)$ at the first order in $g$ for $t$ larger than the bath intrinsic timescales. In the reference frame of the impurity this is given by $\delta \rho(\x,t)=\int \rmd^dy\,\rmd t'\, V(y+\v t',t)\Lambda(\x-\y,t')$ \cite{noauthor_see_nodate}. 
%
Instead of studying the period average $\overline{\delta\rho}$ of $\delta\rho$ (which vanishes), we define: $\delta\kappa(\x)=\overline{\delta\rho(\x,t)\cos t}$. In terms of $\delta\kappa(\x)$, 
the averaged interaction energy is conveniently 
written as
$\overline{H}_{\mathrm{int}}=\int \rmd^d x\,
V_0(\x-\R)\,\delta\kappa(\x),
$
where $V_0(\x)=V(\x,t=0)$. This means that, in the Markovian limit, 
the impurity effectively
interacts with the average wake $\delta\kappa$ via 
the static \emph{repulsive} potential $V_0$.
In the following, we refer to the passive impurity ($\omega=0$) dressed by
its wake $\delta\rho$ as a passive polaron, or simply polaron, and to the
active impurity dressed by $\delta\kappa$ as an \emph{act}ive polar\emph{on}, or \emph{acton} for
short.
The origin of the viscosity and the reversal of its sign become apparent from the structure of the wake. In the passive case, a repulsive polaron moving with velocity $\v$ generates a depletion region that lags behind (relative to the direction of motion) the impurity. 
This is illustrated in  Fig.~\ref{fig:wideimage}(c), which shows $\delta\rho$ for a polaron moving in the positive $x$-direction. The minimum of the wake, marked by the red dashed vertical line, is shifted behind the impurity position (corresponding to the maximum of $V_0$), indicated by the green dashed line. The depleted density exerts an attractive force on the impurity, producing a drag force opposite to the motion and consequently reducing the polaron speed.
In the active case, instead, the average wake $\delta\kappa$ of the acton exhibits multiple depletion and accumulation regions, reflecting the memory of earlier times during which the impurity potential was respectively repulsive and attractive. For the parameter choice corresponding to the profile shown in Fig. \ref{fig:wideimage}(c) (blue line), only the two dominant depletion and accumulation regions overlap significantly with $V_0$ and therefore contribute appreciably to the net force $\F$.
The principal accumulation region is always behind the impurity and thus exerts a force aligned with the direction of motion, propelling the acton forward.
Compared with the passive case, the main depletion region is shifted further forward and, at low velocities, is located ahead of the impurity potential, as visible in the profile $\delta\kappa$  in Fig.~\ref{fig:wideimage}(c). 
%
%
In this regime, the depletion region also contributes to propelling the impurity. As the velocity increases, however, the dominant depletion region progressively shifts to the rear of the impurity. At the self-propulsion velocity, the forces exerted by the depletion and accumulation regions exactly compensate each other, leading to a vanishing net force.
This is illustrated in Fig. \ref{fig:wideimage}(d), which shows $\delta\kappa$ on the $x$--$y$ plane. The center of the principal depletion region is located behind the impurity position, and the force contributions arising from the depletion and accumulation regions are equal and opposite.

\begin{figure}
    \centering
    \includegraphics[width=\linewidth]{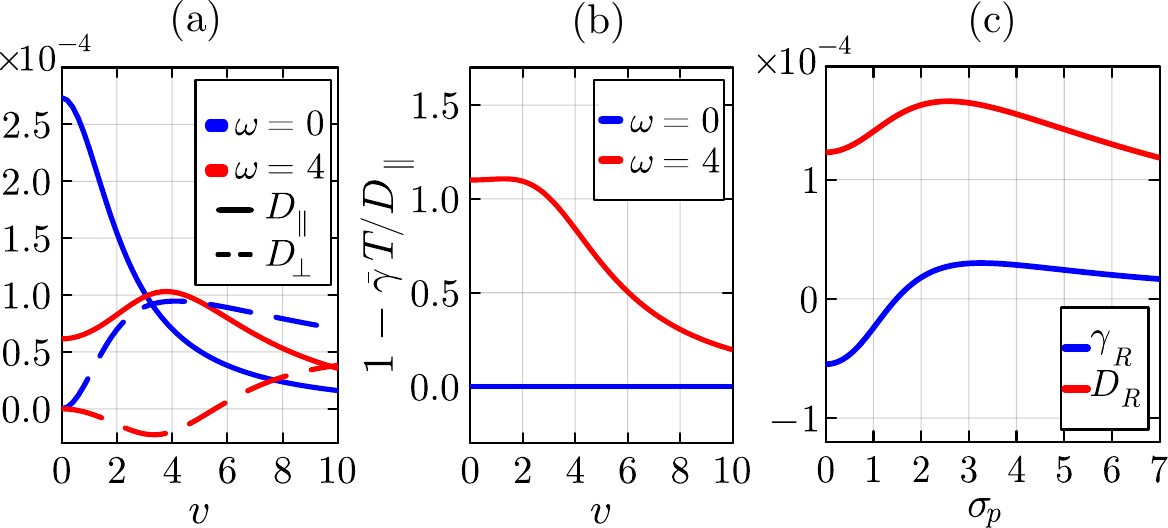}
    \caption{
    (a) Noise strengths $D_{\parallel}$ 
    (solid lines) and $D_{\perp}$ 
    (dashed lines) as functions of $v$ for $\omega = 0$ (blue) and $\omega = 4$ (red). (b) Violation of the FDR as a function of $v$, which vanishes for $\omega = 0$, while it is positive for $\omega = 4$. 
    (c) Plot of $\gamma_R$ (blue line) and $D_R$ (red line) as functions of the variance $\sigma_p$ of the initial state of the particle,
    for $m_I = 1$ and $\omega = 4$. As $\sigma_p$ increases, $\gamma_R$ changes sign from negative to positive, thereby suppressing the self-propulsion observed for $\sigma_p=0$. In all the plots, we set $\hbar=k_B=1$ for $d=3$
    and $m=a=-\mu=g_0=T=1$.
    }
    \label{fig:fluctuations}
\end{figure}

\paragraph*{Fluctuations.}We now consider the role of stochastic and quantum fluctuations, which affect the dynamics both through the effective action $S_A$ and the initial condition $\brho_0$.
In particular, stochastic fluctuations contribute to the action through $S_{\mathrm{fluc}}$, which is quadratic in $\R_q$, while quantum corrections are associated with contributions of order higher than quadratic in $\R_q$, which are neglected in the semiclassical expansion. Indeed, the effective action induced by Eqs.~\eqref{eq:diss_part} and \eqref{eq:fluc_part} is equivalent to the Martin--Siggia--Rose action for a classical particle with position $\R_c$, 
evolving with the Langevin equation
\begin{equation}
\label{eq:langevin}
    m_I \ddot{\R}_c(t)
    =
    \F(\{ \R_c \},t)
    +
    \bm{\zeta}(\R_c(t),t),
\end{equation}
%
%
%
where $\bm{\zeta}(\x,t)$ is a Gaussian random field with correlations
$
    \langle \zeta^i(\x,t)\zeta^j(\x',t') \rangle
    =
    \cD^{ij}(\x-\x',t,t')
$.
Quantum corrections to this dynamics originating from $S_A$ are of order $\hbar^2$, since terms linear in $\hbar$ always vanish~\cite{noauthor_see_nodate}. Accordingly, at first order in $\hbar$, quantum effects enter only via the initial conditions and, in particular, through the constraint that the uncertainty principle imposes on the initial phase-space distribution. 
We will discuss this point in more detail later; for now, we analyze $S_{\rm{fluc}}$ in the Markovian limit by characterizing the mean noise strength $\Bar\DD$. In this regime, we decompose $\Bar\DD^{ij}(\v)=D_{\parallel}(v)\delta^{ij}+D_{\perp}(v)(\delta^{ij}-v^iv^j/v^{2})$, 
where $D_{\parallel}$ and $D_{\perp}$ are plotted in Fig.~\ref{fig:fluctuations}(a) as functions of $v$. 
Note that $D_{\perp}$ may become negative at high $\omega$, while $\Bar\DD$, being a correlation matrix, is positive definite at any $\omega$ and thus its eigenvalues $D_{\parallel}$ and $D_{\parallel}+D_{\perp}$ have to be positive. 
In the passive case $\omega=0$, $D_{\parallel}$ and $D_{\perp}$ satisfy the fluctuation-dissipation relations (FDR): $D_{\parallel}(v)=k_B T\bar\gamma(v)$, $\partial_v(vD_{\perp}(v))=-k_BT\,v\bar\gamma'(v)$ \cite{noauthor_see_nodate}, which are  generically violated at finite $\omega$. 
In Fig.~\ref{fig:fluctuations}(b), we quantify this violation for $D_{\parallel}$
  as a function of $v$, showing that fluctuations exceed dissipation in the active regime. 
  Such excess fluctuations require continuous entropy production \cite{harada_equality_2005} and are therefore a direct signature of active behavior. 
In the presence of self propulsion, the noise $\bm{\zeta}$ causes changes in the direction of the acton.
For small enough temperature $T$, thermal fluctuations of the speed around $v^*$ are negligible and the acton velocity $\v$ in the classical limit can therefore be written as $\v=v^*\hat{\bm{e}}(t)$, where $\hat{\bm{e}}$ is a unit vector pointing in the direction of motion. 
Its time evolution is given by: $\partial_t\hat{\bm{e}}=\sqrt{2D_e}\bm{\eta}(t)$, where $\bm{\eta}(t)$ is a Gaussian noise with variance $\langle\eta^i(t)\eta^j(t')\rangle=(\delta^{ij}-\hat e^i(t)\hat e^j(t))\delta(t-t')$ and $D_e=[D_{\parallel}(v^*)+D_{\perp}(v^*)]/(m_Iv^*)^2$ (see \cite{noauthor_see_nodate}), corresponding to active Brownian motion for the acton.
To study the role of quantum fluctuations, we consider the dynamics of an impurity starting in the fundamental state of an harmonic oscillator of frequency $\Omega$, i.e., in the wavefunction  
$\tilde{\psi}_0(\bm p)=(\sqrt{2\pi}\sigma_p)^{-d/2}e^{-{\bm p}^2/(4\sigma_p^2)}$ 
in momentum space, where $\sigma_p^2=m_I\hbar\Omega/2$ is the momentum uncertainty.
This might be realized by optically trapping the particle.
%
%
Two quantities are relevant for describing the dynamics:
the first is the velocity response function $R_v^{ij}(t,t')=\delta \langle\dot R^i(t)\rangle/\delta F^j(t')$, 
%
i.e., the response of the mean velocity  of the acton at time $t$ to an infinitesimal external force $\delta \F(t)= \delta(t-t')\delta \F$ applied at time $t'$. The second is the symmetrized velocity correlator: $C^{ij}_v(t,t')=\langle\{\dot R^i(t),\dot R^j(t')\}\rangle/2$, where $\{\cdot,\cdot\}$ denotes the anticommutator. 
Both $C^{ij}_v$ and $R^{ij}_v$ 
can be expressed in terms of two-points correlators of $\R_c$ and $\R_q$ which, in turn, we compute by solving the Dyson equation with Eqs.~\eqref{eq:diss_part} and \eqref{eq:fluc_part} as perturbative contributions at order $g^2$ in the Markovian limit \cite{noauthor_see_nodate}. For the response function we find: $R_v^{ij}(t,t')=\delta^{ij}\theta(t-t')\exp\big[-(\gamma_R/m_I)(t-t')\big]/m_I$, where $\theta(t)$ is the Heaviside step function and
\begin{equation}
    \label{eq:gammaR}
    \gamma_R=\frac{m_I^d}{d}\int \rmd^d v \,|\tilde{\psi}_0(m_I \v)|^2\,{\bm \nabla}_v\!\cdot\![\bar\gamma(v)\v].
\end{equation}
%
%
We note that Eq.~\eqref{eq:gammaR} is valid in the limit of large acton mass $m_I$ \cite{noauthor_see_nodate}.
In the classical limit $\hbar = 0$, the renormalized viscosity $\gamma_R$ reduces to $\Bar\gamma(v=0)$ for a static impurity, which may become negative.
However, in the quantum regime, the uncertainty principle does not allow a localized particle to have a well-defined velocity. Consequently, the drag coefficient which is relevant for the acton dynamics is the one averaged 
over the velocity distribution induced by the momentum uncertainty. 
For $\gamma_R<0$, the response function $R_v^{ij}$ grows exponentially in time as the medium enhances the effect of an initial perturbation: the configuration with a static acton is thus unstable, signaling the onset of self propulsion. 
For $\bar{\gamma}(0)<0$, $\gamma_R$ may change sign from negative to positive as $\sigma_p$ increases, indicating that quantum fluctuations can suppress the self-propulsion transition.
This behavior is shown in Fig.~\ref{fig:fluctuations}(c). 
The divergence of $R_v^{ij}(t,t')$ for $t\to\infty$ when $\gamma_R<0$ signals the breakdown of perturbation theory on timescales $t\gtrsim m_I/|\gamma_R|$. 
As a consequence, the present perturbative analysis does not allow us to determine whether the transition is suppressed also at long times. 
Nevertheless, sufficiently precise position measurements can localize the acton within a spatial resolution $\sigma_x$, inducing $\sigma_p=\hbar^2/(2\sigma_x)$ through the uncertainty principle. If the induced momentum fluctuations are large enough to render $\gamma_R>0$, repeated position measurements can dynamically stabilize the static state and suppress the self-propulsion transition over arbitrarily long times. Finally, $C^{ij}_v$ is given by \cite{noauthor_see_nodate}
\begin{equation}
\label{eq:velcorr}
\begin{split}
    C^{ij}_v(t,t')= \left\{\frac{D_R}{\gamma_R}\left[e^{-\frac{\gamma_R}{m_I}|t-t'|}-e^{-\frac{\gamma_R}{m_I}(t+t')}\right]\right.&\\
    \left.\qquad+\frac{\hbar\Omega}{2}e^{-\frac{\gamma_R}{m_I}(t+t')}\right\}\frac{\delta^{ij}}{m_I},&
\end{split}
\end{equation}
with $D_R= (m_I^d/d)\int \rmd^d v\,|\tilde{\psi}_0(m_I \v)|^2\,\mathrm{tr}[\bar\DD(\v)]$. 
In Fig.~\ref{fig:fluctuations}(c), we plot $D_R$ as a function of $\sigma_p$ for an acton in the self-propelling regime. $D_R$ exhibits a non-monotonic dependence on $\sigma_p$, reaching its maximum when $\sigma_p/m_I$ is comparable to the location of the peak in $D_{\parallel}$ (see Fig.~\ref{fig:fluctuations}(a)).
At short times, 
$C^{ij}_v$ is approximately constant, leading to $\langle R^2(t)\rangle=[d\hbar\Omega /(2m_I)]t^2$, i.e., a MSD with ballistic behavior.
This originates from the coherent spreading of the wavefunction under free quantum evolution, which produce a quadratic growth of the MSD at short times. At long times, the dynamics is again controlled by the sign of $\gamma_R$. For $\gamma_R>0$, the exponential decay of the second term on the r.h.s.~of Eq.~\eqref{eq:velcorr} indicates that quantum correlations in the initial state are negligible at long times. The dynamics becomes diffusive, with MSD 
$\langle R^2(t)\rangle=[2d (D_R/\gamma_R^2)] t$. Viceversa, for $\gamma_R>0$ the effect of (both quantum and classical) fluctuations is amplified as the system evolves from the unstable static configuration to the self-propelling regime. In this case, the dynamics at short-times is still described by Eq.~\eqref{eq:velcorr}, which breaks down at longer times, indicating the transition to the self-propelling regime.

\paragraph*{Conclusions.} We have demonstrated that it is possible to induce self-propulsion of polarons (i.e., to generate \emph{act}ive polar\emph{ons} --- \emph{actons}) by periodically modulating the sign of their interaction with the surrounding quantum bath. The transition to this active behavior, described at the semiclassical level by an active Brownian motion, is influenced by the quantum fluctuations and occurs in generic spatial dimensionality.
Here we focused on the emergent semiclassical dynamics, but the phenomenon of activation discussed here should be robust against quantum fluctuations. 
An important direction for future work is indeed the investigation of the effective action beyond leading order in the quantum components $\R_q$. This would allow one to assess the role of quantum effects in the acton dynamics, both within and beyond the Markovian approximation. 
Such an analysis would also clarify the influence of the bath statistics on self-propulsion, which is made inconsequential here by the semiclassical approximation. 
An especially interesting direction for future work, also in view of possible experimental realizations, is to investigate an impurity coupled to a superfluid, where the interplay between self-propulsion and the Landau critical velocity may give rise to novel dynamical phenomena.
Another open question concerns the possibility of velocity-alignment interactions between actons mediated by their wake, potentially of nonreciprocal nature. Such interactions could have important consequences for their collective behavior and may lead to the emergence of collectively traveling states. Finally, we note the possibility to consider systems with a spin-dependent, oscillating bath-impurity coupling, which may give rise to self-propulsion aligned with the spin direction, rather than along a spontaneously selected direction. In light of these considerations, the present work opens multiple new avenues both for understanding and designing active motion at the quantum scales.

\paragraph*{Acknowledgments.} The authors are grateful to Vasco Cavina and Rosario Fazio for valuable and stimulating discussions.

\bibliographystyle{apsrev4-2} 
\bibliography{bibliography.bib}

\onecolumngrid
\newpage

\begin{center}
\textbf{\large Supplementary material for:\\
Self-propulsion of a polaron with oscillating coupling to its quantum bath}\\[1em]
Jacopo Romano  and Andrea Gambassi\\
{\it SISSA --- International School for Advanced Studies and INFN, via Bonomea 265, 34136 Trieste, Italy}
\end{center}
\vspace{1cm}
\setcounter{equation}{0}
\setcounter{figure}{0}
\setcounter{secnumdepth}{3}
\setcounter{page}{1}
\renewcommand{\theequation}{S\arabic{equation}}
\renewcommand{\thefigure}{S\arabic{figure}}
\renewcommand{\thesection}{S.\Roman{section}}

This supplemental material provides details concerning the derivation of the various results presented in the main text.

\section{Derivation of the effective Keldysh action and wake of the impurity}

\begin{figure}[h]
\centering
    \includegraphics[width=0.5\linewidth]{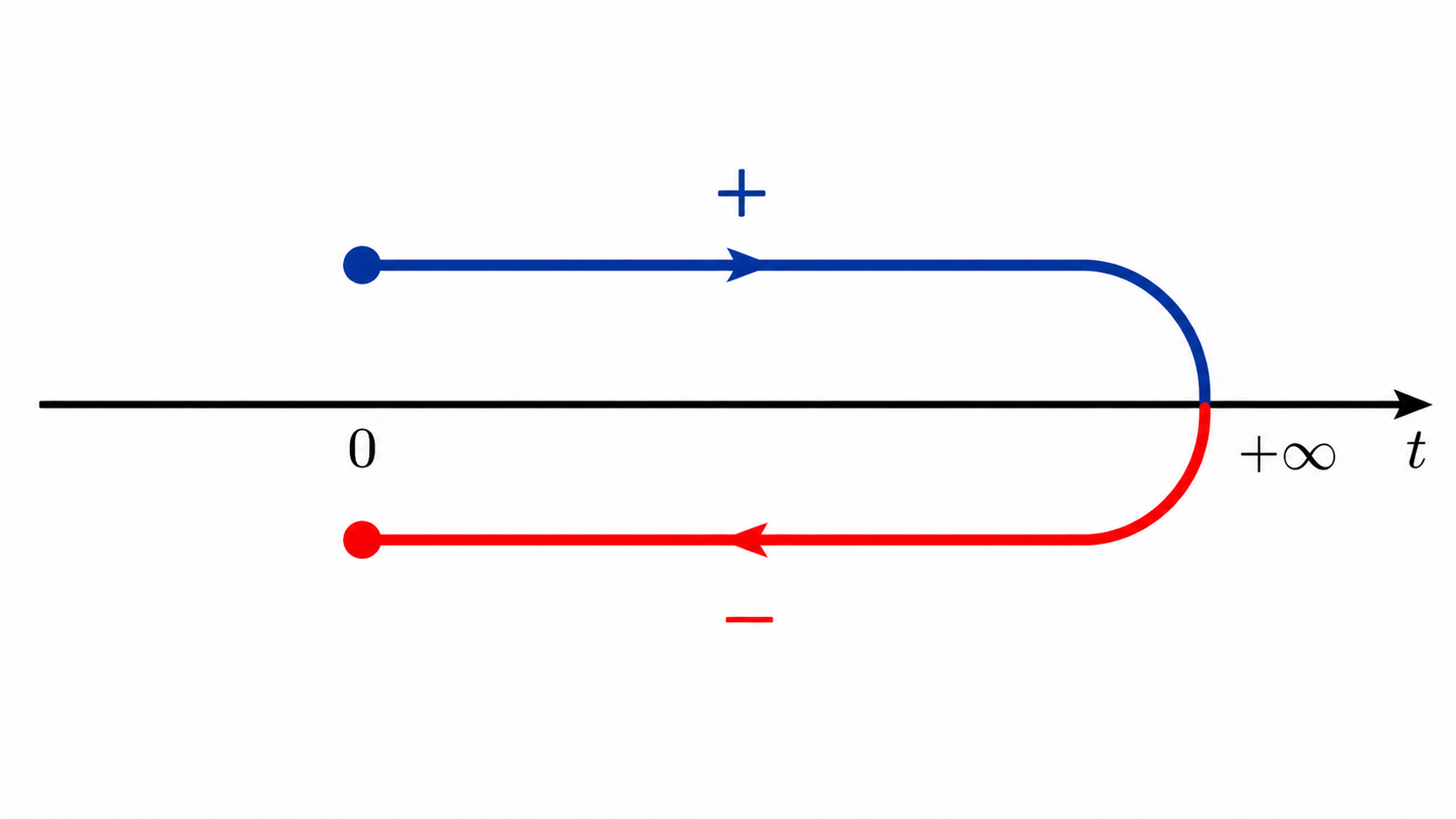}
\caption{Keldysh contour $\mathcal{C}$ running from $0$ to $+\infty$ and back }
    \label{figapp:contour}
\end{figure}

We briefly summarize here the Keldysh formalism used to describe the non-equilibrium dynamics of the impurity coupled to a bosonic bath. The central quantity of this formalism is the generating functional
\begin{equation}
Z = \mathrm{Tr}\!\left[  \,\bm U^\dagger(+\infty,0)\, \bm U(+\infty,0) \brho_0\right],
\end{equation}
where $\bm\rho_0$ is the initial density matrix of the system and $\bm U(t,t')$ its time-evolution operator with $t>t'$.
Although $Z=1$ by construction, it can be turned into a generating function of correlation functions and expectation values by adding suitable sources to its path integral representation, as it is usually done within the functional formalism, see, e.g., Ref.~\cite{kamenev_field_2023} and the discussion below. In fact,
this expression can be represented (see, e.g., Ref.~\cite{kamenev_field_2023}) 
as a path integral over suitable fields of an action $S_{\mathcal{C}}$ in which the time integral is done along the Keldysh contour $\mathcal{C}$, which runs from the initial time $t_0=0$ to $+\infty$ (forward branch, $+$) and back to $0$ (backward branch, $-$), as shown in Fig. \ref{figapp:contour}. In terms of the impurity coordinates $\R(t)$ and the bath field $\phi(\x,t)$, the generating functional reads
\begin{equation}
\label{eqapp:Z}
Z = \int \mathcal{D}[\R,\phi,\phi^*]\;
\langle \R_+^0,\phi_+^0, \phi^{*,0}_+ | \brho_0 | \R_-^0,\phi_-^0 ,\phi^{*,0}_-   \rangle \,
e^{i S_{\mathcal{C}}[\R,\phi,\phi^*]/\hbar},
\end{equation}
%
%
%
where $S_{\mathcal{C}}[\R,\phi,\phi^*]$   is the action of the system constituted by the bath and the impurity, evaluated on the Keldysh contour. The $\pm$ subscripts indicate that the corresponding variables ($\R$, $\phi$, $\phi^*$) are evaluated on the $\pm$ branch of the contour, while the superscript $0$ means that the variable is evaluated at $t=0$. Note that the matrix element of $\bm\rho_0$ couples the two branches at the initial time, encoding the initial conditions.
The action $S_{\mathcal{C}}$ is given by the integral over the Keldish contour of the Lagrangian $\eL(t)$, i.e., $S_{\mathcal{C}}=\int_{\mathcal{C}}{\rm d}t\,\eL(t)$, where $\eL$ can be decomposed as $\eL=\eL_I+\eL_B+\eLi$, with
\begin{eqnarray}
\label{eq:lagrangians}
    &&\eL_B=\int \rmd^dx\,\phi^{*}(\x,t)\lt(i\hbar\partial_t +\frac{\hbar^2}{2m}\nabla^2+ \mu\rt)\phi(\x,t),\nonumber\\
     &&\eL_I =m_I\frac{\dot \R^2(t)}{2}, \quad\mbox{and}\quad \eLi =-\int \rmd^dx \, V(\x-\R(t),t)|\phi(\x,t)|^2. \label{eq:eLint}
\end{eqnarray}
In particular, $\eL_I$ is the Lagrangian associated with the free particle, which can be derived by considering the path-integral representation of the evolution operator determined by the Hamiltonian $H_I$ of the impurity discussed in the main text. Similarly, $\eL_{\phi}$ is derived from the Hamiltonian $H_B$ and 
$\eLi$ from $H_{\rm int}$.
We then perform the Keldysh rotation: $\phi_c=(\phi_++\phi_-)/\sqrt{2}$, $\phi_q=(\phi_+-\phi_-)/(\sqrt{2}\hbar)$, $\R_c=(\R_+ + \R_-)/2$, $\R_q=(\R_+-\R_-)/(2\hbar)$ to write the action as:
\begin{equation}
\label{eqapp:Keldish_action}
 S_{\mathcal{C}}=\int \rmd^dx\,\rmd t \left\{\phi^*_q\lt[i\partial_t+\frac{\nabla^2}{2m}+\mu-V_c(\x,\R,t)\rt]\phi_c+\mbox{c.c.}-V_q(\x,\R,t)(\phi^*_q\phi_q+\phi^*_c\phi_c)\right\}+\int \rmd t \,2 m_I\dot \R_c\dot \R_q,
\end{equation}
where $V_c(\x,\R,t)=[V(\x-(\R_c+\R_q),t)+V(\x-(\R_c-\R_q),t)]/2$ and $V_q(\x,\R,t)=[V(\x-(\R_c+\R_q),t)-V(\x-(\R_c-\R_q),t)]/2$. The argument $\R$ of $V_c$ and $V_q$ indicates collectively both $\R_c$ and $\R_q$.
In order to simplify the notation, in the expression above we have rescaled the reduced Planck constant $\hbar$ and the Boltzmann constant $k_B$ in the various quantites according to
\begin{equation}
\label{eqapp:rescalings}
    m\rightarrow\hbar m, \quad m_I\rightarrow\hbar m_I,\quad  \mu\rightarrow\hbar \mu,\quad V_{c,q}\rightarrow \hbar V_{c,q},\quad T\rightarrow \frac{\hbar T}{k_B} ,\quad \R_q\rightarrow \frac{\R_q}{\hbar}, \quad \mbox{and} \quad \phi_q\rightarrow \frac{\phi_q}{\hbar}.
\end{equation}
The dynamics of the impurity in interaction with the bath can be determined by computing the corresponding effective action $S_A$, obtained by integrating out the field degrees of freedom from $S_{\cal C}$. In particular, one eventually finds $S_A=S_0+S_B$, where $S_0$ is the Keldysh action of the free impurity corresponding to the last term in Eq.~\eqref{eqapp:Keldish_action} and
\begin{equation}
    \label{eqapp:bath_action}
    S_B=-i\log\lt\langle\exp\lt\{-i\int \rmd^dx\, \rmd t \lt[ V_c(\x,\R,t)(\phi^*_q\phi_c+\phi^*_c\phi_q)+V_q(\x,\R,t)(\phi^*_q\phi_q+\phi^*_c\phi_c)\rt]\rt\}\rt\rangle_\phi,
\end{equation}
where the average $\langle \cdots \rangle_\phi$ is taken over the field degrees of freedom.

\subsection{Perturbative calculation of $S_B$}

We compute $S_B$ perturbatively in $V$, organizing the expansion in terms of Feynman diagrams. 

\subsubsection{Propagators}
  The propagators of the free-field theory are indicated by the lines:\\
\begin{center}
\includegraphics[width=.6\linewidth]{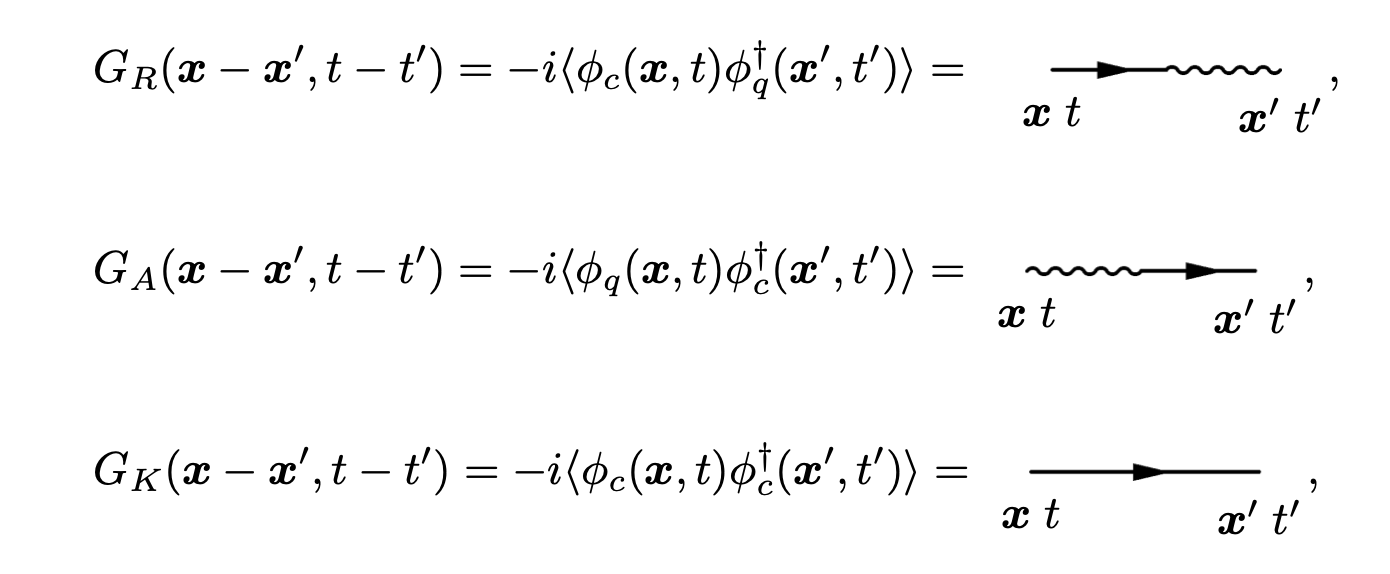}
\end{center}
while $\langle\phi_q(\x,t)\phi^{*}_q(\x',t')\rangle=0$. The Fourier transforms of these propagators are given by:
\begin{equation}
\label{eqapp:propagators}
    G_{R,A}(\k,\en)=\lt(\en-\frac{k^2}{2m}+\mu\pm i0\rt)^{-1}\!\!,\quad\quad G_K(\k,\en)=-2\pi i \coth\lt(\frac{\en}{2T}\rt)\delta\lt(\en-\frac{k^2}{2m}+\mu\rt),
\end{equation}
where $\pm i0$ is an infinitesimal imaginary quantity and its sign is $+$ for $G_R$ and $-$ for $G_A$, while the expression for $G_K$ follows from the fluctuation-dissipation theorem. 
From  Eq.~\eqref{eqapp:propagators} it follows that the inverse Fourier transforms of $G_A$, $G_R$, and $G_K$ are even functions of $\x$ and that
$G_A(\x,t)=G_R^*(\x,-t)$, $G_K(\x,t)=-G_K^*(\x,-t)$. Performing the inverse Fourier transform and inverting the rescaling in Eq.~\eqref{eqapp:rescalings} in order to reinstate the physical units, we obtain
\begin{eqnarray}
    G_R(\x,t)&=&-i \theta(t)\lt(\frac{ m}{2\pi i\hbar t}\rt)^{d/2}\exp\lt(\frac{imx^2}{2\hbar t}+i\frac{\mu t}{\hbar}\rt),
    \\
    G_K(\x,t)&=&- i \int \frac{\rmd^dk}{(2\pi)^{d}} \, \coth\lt(\frac{\frac{\hbar^2k^2}{2m}-\mu}{2k_B T}\rt)\exp\lt[i\k\cdot\x -i\lt(\frac{\hbar k^2}{2m}-\frac{\mu}{\hbar}\rt)t\rt].
\end{eqnarray}
In the expression for $G_R$, $\theta(t)$ is the step function with $\theta(t> 0) = 1$ 
and  $\theta(t< 0) = 0$, while the power $d/2$ of the prefactor is defined by taking  first the square root (on the principal branch) and then raising the result to the $d$-th power, thereby avoiding branch cuts. The same convention applies throughout.
As discussed in the main text, we focus here on the semiclassical limit in order to show the emergence of activity. In this limit, the thermal energy $k_BT$ is larger than both the chemical potential $\mu$ and the kinetic energy $E_K$ of the wake (as defined in the main text). Accordingly, one can expand the cotangent in $G_K$ for small argument and, using the identity $A^{-1}=\int_0^{\infty}\rmd z\, e^{-Az}$, one can eventually write
\begin{equation}
    G_K(\x,t)=-2i\frac{k_B T}{\hbar}\lt(\frac{m}{2\pi\hbar}\rt)^{d/2}\int_0^{\infty} \rmd z \frac{1}{(z+it)^{
d/2}}\exp\lt[-\frac{mx^2}{2\hbar(z+it)}+\frac{\mu}{\hbar} z\rt].
\end{equation}

\subsubsection{Vertices and Feynmann diagrams}
\begin{figure}
    \centering
    \includegraphics[width=0.75\linewidth]{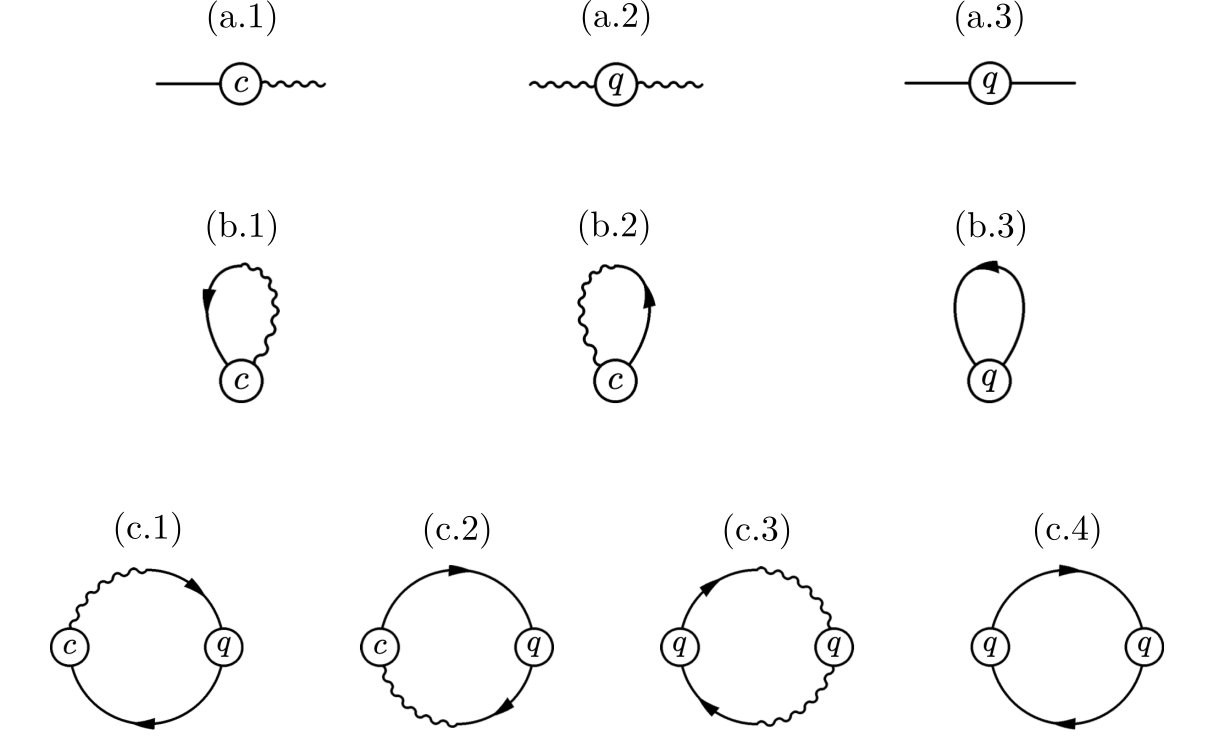}
    \caption{(a.1)--(a.3) Vertices, (b.1)--(b.3) and (c.1)--(c.4) diagrams for the perturbative calculation of $S_B$.}
    \label{fig:diagrams}
\end{figure}

The interaction vertices generated by Eq.~\eqref{eqapp:bath_action} can be represented as in Fig.~\ref{fig:diagrams}(a.1)--(a.3),
where the vertex in (a.1) is associated to $V_c$ while those in (a.2) and (a.3) are associated to $V_q$. Accordingly, when a free propagator is connected to one of this vertices (e.g., at its  $\x,t$ endpoint) it is multiplied by the corresponding potential $V_{c/q}(\x,t)$ and the result is integrated in $\x$ over all space. 
At first order in $V$, the corresponding diagrams are shown in Fig.~\ref{fig:diagrams}(b.1)--(b.3).
Their contribution $S_B^{(1)}$ to $S_B$ is (in rescaled units) given by
\begin{equation}
    \label{eqapp:S1}
    S_B^{(1)}= \int \rmd t \int \rmd^dx \,\left\{ V_c(\x,\R)\left[G_R(\mathbf{0},0)+G_A(\mathbf{0},0)\right]+V_q(\x,\R)G_K(\mathbf{0},0)\right\},
\end{equation}
which vanishes because $G_R(\mathbf{0},0)+G_A(\mathbf{0},0) = 0$
and $\int \rmd^dx\, V_q(\x,t)=0$. The fact that the contribution in Eq.~\eqref{eqapp:S1} vanishes is actually a consequence of the normalization condition $Z=1$ implied by Eq.~\eqref{eqapp:bath_action}, which requires that the contribution of vacuum diagrams, such as those in Fig.~\ref{fig:diagrams}(b.1)--(b.3), vanish identically (as otherwise they would modify the value of $Z$) \cite{kamenev_field_2023}.
At the second order the (nonvanishing) diagrams are shown in Fig.~\ref{fig:diagrams}(c.1)--(c.4).
These give the second order $S^{(2)}_B$ contribution to the action:
\begin{align}
\label{eq:S2}
    S_B^{(2)}&=-i\int \rmd^dx\rmd^dx' \rmd t\, \rmd t' \Big\{V_q(\x,\R(t),t)V_c(\x',\R(t'),t')[G_R(\Delta\x,\Delta t)G_K(-\Delta\x,-\Delta t)\nonumber\\
    &\hspace{9.5cm}+G_K(\Delta\x,\Delta t)G_A(-\Delta\x,-\Delta t)]\nonumber\\&+V_q(\x,\R(t),t)V_q(\x',\R(t'),t')\lt[\frac{1}{2}G_K(\Delta\x,\Delta t)G_K(-\Delta\x,-\Delta t)+G_R(\Delta\x,\Delta t)G_A(-\Delta\x,-\Delta t)\rt]\Big\},
\end{align}
where $\Delta\x=\x-\x'$ and $\Delta t=t-t'$.
The resulting action $S_B = S_B^{(2)}=S_{\rm diss}+S_{\rm fluc}$ 
contains two contributions: $S_{\rm diss}$ describes dissipation, while $S_{\rm fluc}$ encodes thermal fluctuations. 
Using the aforementioned identities of the Green's functions, we find that these contributions are respectively given by 
\begin{equation}
    \label{eq:dissact1}
    S_{\rm{diss}}=2\int \rmd^dx\rmd^dx' \rmd t \,\rmd t'\,  V_c(\x',\R(t'),t')V_q(\x,\R(t),t) \,\mathrm{Im}\left[G_R(\Delta\x,\Delta t)G_K(-\Delta \x,-\Delta t)\right],
\end{equation}
and
\begin{equation}
    \label{eq:fluctact1}
    S_{\rm{fluc}}=i\int \rmd^dx\rmd^dx' \rmd t \,\rmd t' \, V_q(\x',\R(t'),t')V_q(\x,\R(t),t) \lt[\frac{1}{2}|G_K(\Delta \x,\Delta t)|^2-|G_R(\Delta\x,\Delta t)|^2\rt].
\end{equation}
We now take the semiclassical limit, expanding the effective action of the particle up to the second order in the quantum component $\R_q$ of the position of the impurity. Expanding the potentials one has $V_c(\x,\R,t)\simeq V(\x-\R_c,t)$ while $V_q(\x,\R,t)\simeq -\R_q \cdot {\bm \nabla} V(\x-\R_c,t)$. We also note that the term in $S_{\rm{fluc}}$ containing $|G_R(\Delta\x,\Delta t)|^2$ originates from a correlator containing two quantum components of the field and therefore it leads to a contribution of higher order in $\hbar$ which is subleading compared to the term containing $|G_K(\Delta\x,\Delta t)|^2$.
Combining all the leading 
contributions in the semiclassical expansion, we obtain:
\begin{equation}
    \label{eq:sb1}
    S_A=S_0+  2 \int\rmd t\, \R_q(t)\cdot\int^t_{-\infty}\rmd t'\E(\Delta\R_c,t,t')+ 2i\int \rmd t\, \rmd t'\, \R_q(t')\cD(\Delta\R_c,t,t')\R_q(t),
\end{equation}
where $\Delta \R_c=\R_c(t)-\R_c(t')$ and
\begin{align}
    \label{eq:dissipation}
    E^i(\x,t,t')&=\partial^i \int \rmd^d y\,\rmd^d Y\, V(\Y+\y/2,t)V(\Y-\y/2,t')\mathrm{Im}\left[G_R(\x+\y,\Delta t)G_K(\x+\y,-\Delta t)\right], \\
    \label{eq:fluctuation}
    \cD^{ij}(\x,t,t')&=-\partial^i\partial^j\int \rmd^d y\,\rmd^d Y\, V(\Y+\y/2,t)V(\Y-\y/2,t')\frac{1}{4}|G_K(\x+\y,t-t')|^2.
\end{align}
The effect of the interaction on the classical field $\delta\rho(\x,t)$ 
is given by the expectation value of $|\phi_c(\x,t)|^2$ 
calculated along a fixed classical trajectory with $\R_c=\v t$ and $\R_q=0$. 
Expanding the action in Eq.~\eqref{eqapp:Keldish_action} at first order in $V \propto g$ we obtain:
\begin{equation}
\label{eqapp:classfield}
    \delta\rho(\x,t)=-i\int \rmd^dy\rmd t'\, V(\y-\R(t'),t')\langle \phi_c(\x,t)\phi^{*}_c(\x,t)[\phi^{*}_c(\y,t')\phi_q(\y,t')+\phi_c(\y,t')\phi^{*}_q(\y,t')]\rangle_{\phi}.
\end{equation}
Using Wick's theorem and the definition of $G_R$, $G_R$ and $G_K$ we obtain the result reported in the main text.

\section{Fluctuation-dissipation relation at $\omega=0$}

The non-Markovian fluctuation-dissipation relation (FDR) $\partial_t \cD^{ij}(\x,t-t')=k_BT\partial^iE^j(\x,t-t')$ reported in the main text can be derived from the identity 
\begin{equation}
\label{eqapp:partial_tGk}
    \partial_t G_K(\x,t)=\frac{2ik_BT}{\hbar}[G_R^*(\x,t)-G_R^*(\x,-t)],
\end{equation} 
which is nothing but the expression of the (classical) fluctuation-dissipation theorem for the correlation functions of a bath in equilibrium at high temperature $T$, see, e.g., Ref.~\cite{kamenev_field_2023}. 
In turm, this equality follows from expanding Eq.~\eqref{eqapp:propagators} at high $T$, deriving w.r.t.~$t$ and then calculating the inverse Fourier transform. 
Then, deriving $\cD$ in Eq.~\eqref{eq:fluc_part} w.r.t.~$t$ and using Eq.~\eqref{eqapp:partial_tGk}, we obtain the desired FDR relation. To obtain its Markovian limit we note that, at $\omega=0$, one has  $\DD(\x,t,t')=\DD(\x,t-t')$ and consider the identity
\begin{equation}
\label{eqapp:markFDTstep1}
    \frac{\rmd }{\rmd t}\cD^{ij}(\v t,t)=\frac{\partial }{\partial t}\cD^{ij}(\v t,t)+\frac{v^k}{t}\frac{\partial }{\partial v^k}\cD^{ij}(\v t,t)
\end{equation}
for a particle  moving with velocity $\v$. Integrating both sides in $t$ from $-\infty$ to $0$ and using the non-Markovian FDR on $\partial_t \cD^{ij}(\v t, t)$ we obtain the relation:
\begin{equation}
    \label{eqapp:markFDR}
    \DD^{ij}(\v)+v^k\frac{\partial }{\partial v^k}\DD^{ij}(\v)=\frac{\partial}{\partial v^i} [\gamma(v)v^k].
\end{equation}
Using the decomposition $\DD^{ij}(\v)=D_{\parallel}(v)\delta^{ij}+D_{\perp}(v)(\delta^{ij}-v^iv^j/v^{2})$ in Eq.~\eqref{eqapp:markFDR}, we obtain the relations between $D_{\parallel}$, $D_{\perp}$, and $\gamma$ reported in the main text.

\section{Reduction of the classical dynamics to an active Brownian motion}
In the main text, we argued that, in the Markovian limit and for sufficiently weak thermal fluctuations, the dynamics described by Eq.~\eqref{eq:langevin} reduces to that of an active Brownian particle, and we reported the corresponding effective angular diffusion coefficient. To derive this result, we consider Eq.~\eqref{eq:langevin} in the Markovian limit and assume that the temperature  $T$ is sufficiently low that the speed $v$ remains close to the value $v^*$ of self-propulsion. 
We then expand 
$\bar\gamma$ and $\bar\DD$ for $v$ about $v^*$, retaining the leading non-vanishing terms, which yields
\begin{equation}
    \label{eqapp:expanded_langevin}
    m_I\dot \v=-\bar\gamma'(v^*)(v-v^*)\v +\bm{\zeta}(t),
\end{equation}
with $\langle\zeta^i(t)\zeta^j(t')\rangle=\bar{\DD}^{ij}(v^*)\delta(t-t')$. 
Indicating by $\hat{\bm e}(t)$ the unit vector aligned with $\v(t)$, we project the dynamics in 
Eq.~\eqref{eqapp:expanded_langevin} on its longitudinal and transverse components by multiplying  both sides of Eq.~\eqref{eqapp:expanded_langevin} by $\hat{\bm e}\,\cdot$ and $-\hat{\bm e}\times\hat{\bm e}\times$, respectively. 
Since the longitudinal component is damped by the friction, oscillations of $v$ around $v^*$ are negligible at sufficiently low $T$. For the longitudinal component, instead, we have:
\begin{equation}
\label{eq:}
    m_I v^* \partial_t  \hat e^i(t)=[\delta^{ij}-\hat e^i(t)\hat e^j(t)]\zeta^j(t).
\end{equation}
Defining $\eta^i(t)=1/(m_I v^*)[\delta^{ij}-\hat e^i(t)\hat e^j(t)]\zeta^i(t)$ and computing its correlator we obtain the dynamics reported in the main text.

%
%

\section{Perturbative self-energy of the acton and Dyson equation}

The Green function $\bm \G$  for the acton dynamics is a matrix-valued function with component: $\G_{\alpha\beta}^{ij}(t_1,t)=\frac{i}{2}\langle R_\alpha^i(t_1)R_\beta^j(t)\rangle$, where the indices $\alpha,\,\beta \in \{c,\, q\}$ refer to the  quantum or classical components while the  indices $i$ and $j$ indicate the spatial components. The time evolution of $\G_{\alpha\beta}^{ij}$ is described by the Dyson equation in terms of the unperturbed Green function $\bm \G_0$ for a free particle (i.e., with $g=0$) and the self energy $\bm \Sigma$. The latter is defined as the sum of all the irreducible diagrams that can be obtained by connecting the vertices induced by $S_B=S_{\rm{diss}}+S_{\rm{fluc}}$ with the free propagator $\bm\G_0$.  In particular, the Dyson equation reads: $[\bm{\G}^{-1}_0-\bm{\Sigma}]\,\circ\, \bm{\G}=\mathds{1}$, where $\circ$ denotes matrix product in the spatial and classical/quantum components and convolution in time, while the identity is given by $\mathds{1}^{ij}_{\alpha\beta}(t,t')=\delta^{ij}\delta_{\alpha\beta}\delta(t-t')$. The components of the inverse $\bm{\G}^{-1}_0$ of the free propagator are, instead, given by $(\G^{-1}_{0})_{\alpha\beta}^{ij}(t_1,t)=-m_I\sigma_{\alpha\beta}\delta^{ij}\partial_{t_1}^2\delta(t_1-t)$, where we introduced the matrix
\begin{equation}
    \bm{\sigma}=\begin{pmatrix} 0 & 1 
    \\ 1 & 0 
    \end{pmatrix}.
\end{equation}
We compute the self-energy $\bm{\Sigma}$ to first order in perturbation theory in $S_B$. At this order, diagrams which contribute to the expansion contain a single interaction vertex and are therefore irreducible. Consequently, the self-energy can be written as:
\begin{eqnarray}
    \label{eqapp:self_energy}
    \Sigma_{\alpha\beta}^{ij}(t_1,t_2)=-\frac{1}{2}\left\langle\frac{\delta^2S_B/\hbar}{\delta R^i_\alpha(t_1)\delta R^j_\beta(t_2)}\right\rangle_0,
\end{eqnarray}
where $\langle\cdots\rangle_0$ indicates that the expectation value is computed in the free theory. To evaluate the expression in Eq.~\eqref{eqapp:self_energy} we start by noticing that causality implies $\Sigma_{cc}^{ij}(t_1,t_2)=0$ and $\Sigma_{cq}^{ij}(t_1,t_2)=\left[\Sigma_{qc}^{ij}(t_2,t_1)\right]^*$.
In order to determine $\bm{\Sigma}_{qq}$ in the Markovian limit from Eq.~\eqref{eqapp:self_energy}, we note that the part of $S_B$ which is quadratic in $\R_q$ is the fluctuation action $S_{\rm fluc}$ in Eq.~\eqref{eq:fluc_part}, i.e.,  
\begin{equation}
    \label{eqapp:mark_fluc}
    S_{\rm{fluc}}=4i\hbar\int \rmd t\, R_q^i(t)\,R_q^j(t) \,\overline{\DD}^{ij}[\dot R_c(t)].
\end{equation}
Calculating the functional derivative of this expression with respect to $\R_q(t)$ and $\R_q(t')$ we find
\begin{equation}
    \label{eqapp:sigmaqq}
    \bm{\Sigma}_{qq}(t_1,t_2)=-4i\langle \overline{\DD}[\dot R_c(t_1)]\rangle_0\delta(t_1-t_2).
\end{equation}
To compute the expectation value in Eq.~\eqref{eqapp:sigmaqq} we notice that, since the free action is quadratic and the initial condition is gaussian, $\dot R_c^i(t)$ is gaussian at any time $t$. Moreover, since  momentum is conserved in the free theory, the variance of $\dot R^i_c$ is constant and it is related to the variance of the momentum by $\langle\dot R^i_c(t)^2\rangle =\sigma_p^2/m_I^2$. 
Finally, since the initial momentum distribution is isotropic, $\Sigma^{ij}_{qq}$ is also isotropic and we can write that $\Sigma^{ij}_{qq}(t_1,t_2)=-4i D_R\delta(t_1-t_2)\delta^{ij}$, where $D_R$ is defined as below Eq.~\eqref{eq:velcorr} in the main text.
Concerning the determination of $\bm{\Sigma}_{cq}$, we fist decompose it as $\bm{\Sigma}_{cq}=\bm{\Sigma}_{cq,{\rm diss}}+\bm{\Sigma}_{cq,{\rm fluc}}$, where the  contributions $\bm{\Sigma}_{cq,{\rm diss}}$ and $\bm{\Sigma}_{cq,{\rm fluc}}$ are given, respectively, by the dissipative $S_{\rm diss}$ and the fluctuation part $S_{\rm fluc}$ of the action $S_B$. The dissipative contribution can be computed directly in the Markovian limit, for which $S_{\rm diss}$ is given by
\begin{equation}
    S_{\rm{diss}}=-2\int_0^{\infty}dt \,R^i_q(t) \bar{\gamma}(\dot R_c(t))\,\dot R_c^i(t).
\end{equation}
Then, using the fact that, for a generic functional $G[\dot R]$, one has $\delta G/\delta R^i(t)=-\partial_t[\delta G/\delta \dot R^i(t)]$, 
we obtain:
\begin{equation}
\label{eqapp:sigmadiss1}
    \Sigma^{ij}_{qc,{\rm diss}}(t_1,t_2)=-\left\langle \frac{\partial [\dot R_c^i(t)\bar{\gamma}(\dot R_c(t))]}{\partial \dot R_c^j(t)}\right\rangle_0\partial_{t_2}\delta(t_1-t_2)\delta^{ij}.
\end{equation}
The expectation value in this expression can be computed with similar consideration as for $\langle \bar\DD(\dot R_c)\rangle_0$ in Eq.~\eqref{eqapp:sigmaqq}. This leads to $\Sigma^{ij}_{qc,{\rm diss}}(t_1,t_2)=-\gamma_R\partial_{t_2}\delta(t_1-t_2)$, with $\gamma_R$ given by Eq.~\eqref{eq:gammaR} of the main text. 
The determination of $\bm{\Sigma}_{qc,{\rm fluc}}$ in the Markovian limit requires more care, since a direct calculation based on Eq.~\eqref{eqapp:mark_fluc} would provide a result proportional to the ill-defined value $\theta(0)$ of the step function.
In order to resolve avoid this ambiguity, it is convenient to compute $\bm{\Sigma}_{qc,{\rm fluc}}$ from the non-Markovian action in Eq.~\eqref{eq:fluc_part} and take the Markovian limit only at the end of the calculation. 
Taking the functional derivative of Eq.~\eqref{eq:fluc_part} w.r.t.~$R_q^i(t_1)$ and $R_c^j(t_2)$, in fact, we find
\begin{align}
    \label{eqapp:nonmark}
    \Sigma_{qc,{\rm fluc}}^{ij}(t_1,t_2)=2i\bigg[&\langle R_c^l(t_1)R^k_q(t_2)\rangle_0\langle\nabla^j\nabla^k\cD^{ik}(\R_c(t_2)-\R_c(t_1),t_2,t_1)\rangle_0\\&\nonumber-\int \rmd t\langle R_c^l(t)R^k_q(t_2)\rangle_0\langle\nabla^j\nabla^k\cD^{ik}(\R_c(t_2)-\R_c(t),t_2,t)\rangle_0\,\delta(t_1-t_2)\bigg],
\end{align}
where we use Wick theorem to decompose the expectation value in Eq.~\eqref{eqapp:self_energy} as the product of two expectation values. 
In the Markovian limit, we approximate $\R_c(t_2)-\R_c(t_1)\simeq \dot\R_c(t_2) (t_1-t_2) $ and we consider only the average of Eq.~\eqref{eqapp:nonmark} w.r.t.~the oscillations of the coupling $g(t)$. We then apply $\Sigma_{qc,{\rm fluc}}^{ij}(t_1,t_2)$ on a test function $\mathcal{T}(t)$. Since $\cD$ decays rapidly upon increasing $t_1-t_2$,  we can approximate $\mathcal{T}(t)=\mathcal{T}(t_1)+\mathcal{T}'(t_1)(t_1-t_2)$. The term $\propto g(t_1)$ vanishes and we eventually obtain:
\begin{equation}
    \label{eqapp:Dg}
    \int \rmd t_2 \, \Sigma_{qc,{\rm fluc}}^{ij}(t_1,t_2)\mathcal{T}(t_2)=-\frac{\mathcal{T}'(t_1)}{m_I}\int_{-\infty}^{t_1}\rmd t_2 \, (t_1-t_2)^2\langle\nabla^j\nabla^k\bar{\cD}^{ik}(\dot\R_c(t_2)(t_1-t_2),t_1-t_2)\rangle_0,
\end{equation}
where we used the fact that $\langle R_c^i(t)R^j_q(t')\rangle_0=i/(2m_I) (t'-t)\theta(t-t')$.
Finally, noting that $(t_1-t_2)\nabla^i=\partial/\partial \dot{R}_c^i(t_2)$ we obtain
\begin{equation}
    \label{eqapp:sigmaqcfluc}
    \Sigma^{ij}_{qc,{\rm fluc}}(t_1,t_2)=\frac{1}{m_I}\left\langle\frac{\partial^2 \overline{\DD}^{ik}}{\partial v^j\partial v^k}\right\rangle_0\partial_{t_2}\delta(t_2-t_1)
\end{equation}
%
Note that this term gives rise to a stochastic drift contribution to the probability current, originating from the multiplicative nature of the noise, which adds to the deterministic drift in Eq.~\eqref{eqapp:sigmadiss1}. However, this contribution becomes negligible in the limit of sufficiently large $m_I$. For simplicity, we assume throughout that this condition is satisfied.
Having determined the self-energy, we can compute the velocity response and correlation functions, $R_v^{ij}$ and $C_v^{ij}$, from the Dyson equation.
The response function is defined as
 $R^{ij}_v(t,t')=-\partial_t\G_{cq}^{ij}(t,t')$. Using the Dyson equation and Eq.~\eqref{eqapp:sigmadiss1} we find that $R^{ij}_v$ satisfies the differential equation:
\begin{equation}
\label{eqapp:Rv}
    m_I\partial_t R^{ij}_v(t,t')+\gamma_R R^{ij}_v(t,t')=\delta^{ij}\delta(t-t').
\end{equation}
Solving Eq.~\eqref{eqapp:Rv} and imposing that $R_v(t,t')=0$ for $t<t'$ leads to the result reported in the main text just above Eq.~\eqref{eq:gammaR}.
The correlation function $C^{ij}_v$ of the velocities is, instead, given by $C^{ij}_v=i/2\,\,\partial_t\partial_{t'}\G_{cc}(t,t')$. Using the Dyson equation, Eqs.~\eqref{eqapp:sigmadiss1} and \eqref{eqapp:sigmaqq} we find that $C^{ij}_v$ satisfies the equation:
\begin{equation}
\label{eqapp:Cv}
    m_I\partial_t C^{ij}_v(t,t')+\gamma_R C^{ij}_v(t,t')=2D_R \delta^{ij}e^{\frac{\gamma_R}{m_I}(t-t')}\theta(t'-t).
\end{equation}
Solving this equation with the initial condition $C^{ij}_v(0,0)=\sigma_p^2/m_I^2\delta^{ij}$ and imposing $C_v^{ij}(t,t')=C_v^{ij}(t',t)$ we obtain Eq.~\eqref{eq:velcorr} of the main text.

\end{document}